\documentclass[aps,prb,preprint,showpacs,superscriptaddress]{revtex4-1}
\usepackage{graphicx}
\begin{document}
\title{Magnonic band structure of domain wall magnonic crystals}
\author{D. Wang}
\email{daowei\_wang@csu.edu.cn}
\affiliation{School of Physics and Electronics, Central South University, Changsha 410083, Hunan, P. R. China}
\author{Yan Zhou}
\affiliation{Department of Physics, The University of Hong Kong, Hong Kong, P. R. China}
\author{Zhi-xiong Li}
\affiliation{School of Physics and Electronics, Central South University, Changsha 410083, Hunan, P. R. China}
\author{Yaozhuang Nie}
\affiliation{School of Physics and Electronics, Central South University, Changsha 410083, Hunan, P. R. China}
\author{Xi-guang Wang}
\affiliation{School of Physics and Electronics, Central South University, Changsha 410083, Hunan, P. R. China}
\author{Guang-hua Guo}
\email{guogh@csu.edu.cn}
\affiliation{School of Physics and Electronics, Central South University, Changsha 410083, Hunan, P. R. China}
\begin{abstract}
Magnonic crystals are prototype magnetic metamaterials designed for the control of spin wave propagation. Conventional magnonic crystals are composed of single domain elements. If magnetization textures, such as domain walls, vortices and skyrmions, are included in the building blocks of magnonic crystals, additional degrees of freedom over the control of the magnonic band structure can be achieved. We theoretically investigate the influence of domain walls on the spin wave propagation and the corresponding magnonic band structure. It is found that the rotation of magnetization inside a domain wall introduces a geometric vector potential for the spin wave excitation. The corresponding Berry phase has quantized value $4 n_w \pi$, where $n_w$ is the winding number of the domain wall. Due to the topological vector potential, the magnonic band structure of magnonic crystals with domain walls as comprising elements differs significantly from an identical magnonic crystal composed of only magnetic domains. This difference can be utilized to realize dynamic reconfiguration of magnonic band structure by a sole nucleation or annihilation of domain walls in magnonic crystals.

\textit{Index Terms} - Domain walls, spin waves, magnonic bands
\end{abstract}
\date{\today}
\maketitle
\section{Introduction}
Spin waves (SWs) are fundamental elementary excitations in magnetically ordered solid systems. Originally, the concept of SW was proposed to explain the famous $T^{3/2}$ law of the temperature dependence of saturation magnetization. Thereafter, the existence of SWs was confirmed by many experiments, and SWs became a basic ingredient in the field of magnetism and magnetic materials. Thanks to the rapid development of micro-structuring technology, even the manipulation of the propagation of SW itself in periodic magnetic structures, dubbed magnonic crystals (MCs) \cite{magnonics1,magnonics2}, becomes feasible nowadays. In practice, this degree of freedom in the manipulation of SW dynamics paves the route to information processing employing SWs. However, most of the contemporary studies involve only patterned magnetic domains and antiferromagnetically coupled nanowires \cite{Topp10}, the role of magnetization texture still awaits for investigation \cite{texturemag,Li15}. The inclusion of topological magnetization textures into the building elements of MCs would not only enlarge the horizon of the quest for new types of MCs, but also benefit from the additional tunability coming along with the response of magnetization textures to externally applied magnetic field \cite{reconfig} or electric current \cite{skyrmion}. Actually, the interplay between topology and SW has already attracted theoretical interest, both in MCs \cite{Shindou13} and in topological magnon insulators \cite{Mook14}.

Despite of the superficial different behaviour exhibited by DWs and SWs, they are closely related to each other: SWs are the propagating excitation of a ferromagnet, while the wavefunction of the zero mode excitation of a ferromagnet is just the derivative of the DW profile \cite{braun,Tatara08}. Here, we will consider the 1D band structure where there is a 2$\pi$ domain wall (DW), which is the simplest topological entity in magnetic materials \cite{braun}, in the unit cell of a MC. By analytically solving the SW eigenequation, we demonstrate that the topological nature of the underlying DW is transferred to the Berry phase experienced by the SW. Due to this topological nature of the Berry phase, the vector potential giving rise to the Berry phase determines a different magnonic band structure, compared to the band structure of an identical conventional MC composed of only uniform domains. We believe that the same topological effect is also effective in the skyrmion-based dynamic MCs \cite{skyrmion}, given that the homotopy group of both 2$\pi$ DWs and skyrmions is identical \cite{braun}.

\begin{figure}\centering
\begin{minipage}[c]{0.6\linewidth}
\includegraphics[width=\linewidth]{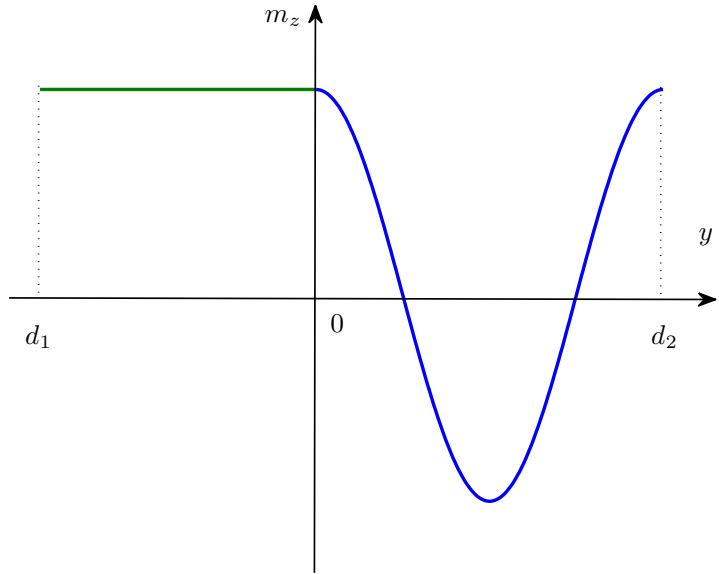}
\end{minipage}
\caption{Schematic distribution of the $z$ component of the magnetization vector, $m_z$, in the unit cell of 1D DWMC. Region 1 (uniform domain region) is from $-d_1$ to 0, and region 2 (DW region) from 0 to $d_2$. For the case of a conventional domain MC, the two region are occupied by two different magnetic materials and the 2$\pi$ DW shown here is absent. For a DWMC, materials in the two regions could be identical or different. Although the DW shown here is a 2$\pi$ DW, $\pi$ DWs can also be included in a DWMC. The corresponding unit cell for a DWMC with $\pi$ DWs is more complicated than the unit cell shown here.}
\label{cell}
\end{figure}

The fundamental unit cell of a domain-wall magnonic crystal (DWMC) is shown schematically in Fig. \ref{cell}. It is composed of two regions, region 1 and region 2. In our coordinate system, region 1 extends from $y = - d_1$ to $y = 0$, and region 2 is bounded by $y = 0$ and $y = d_2$ planes. The magnetic materials in the two regions could be identical or different. For a usual MC, in order to induce an energy gap for SWs, the magnetic parameters of the two regions have to be different, which can be realized by using different materials or local modification of the magnetic properties of the same material through ion implantation \cite{Franken12,Buckingham12}. In our simplified 1D treatment of DWMCs, what matters is the magnetic parameters in the two regions, which are differentiated by the subscripts 1 and 2 for regions 1 and 2, respectively. For concreteness of the problem, we assume both regions have uniaxial perpendicular magnetic anisotropy along the $z$-axis and the MC structure is patterned from a continuous film lying in the $xy$ plane.

The organization of our discussion of DWMCs is as follows. In Sec. \ref{sws}, we give the SW eigenfunctions by solving the linearized magnetization dynamics. With those magnonic eigenfunctions, the band structure of MCs composed of both single domains and domain walls inside the unit cell is studied in Sec. \ref{band}, by linearly superposing the eigenfunctions in each region. In Sec. \ref{gap}, the dependence of the magnonic band gaps and widths on magnetic material parameters will be investigated. Using the method of transfer matrix, the evolution from free propagating SWs without band gaps to SWs characterized by band structure is described in Sec. \ref{evol}. Finally, Sec. \ref{discussion} discusses the possibility of creating DWMCs in experiments.

\section{Spin wave eigenmodes}
\label{sws}
Before proceeding to the actual band structure of the MC, we first need to know the SW eigenmodes in each material. This requires the solution of the corresponding Landau-Lifshitz-Gilbert (LLG) equation \cite{llg}. Without an externally applied magnetic field and neglecting the damping term, the LLG equation for the normalized magnetization vector $\textbf {m} = \textbf {M}/M_s$ reduces to
\begin{equation}
\label{llg}
- \frac{d \textbf {m}}{dt} = \omega_c \textbf {m} \times (m_z \hat {z} + \delta^2 \nabla ^2 \textbf {m}),
\end{equation}
with $\omega_c = \gamma H_K$ the cutoff frequency for SWs and $\delta = \sqrt {A/K}$ the DW width constant. $A$ is the exchange stiffness constant, and $K$ the uniaxial anisotropy constant. $H_K$ is the anisotropy field, $H_K = 2 K/M_s$, with $M_s$ the saturation magnetization. \textit{In Eq. (\ref{llg}), we consider only a perpendicular anisotropy field. In our simplified 1D treatment of the SW dynamics, the demagnetization energy simplifies to two uniaxial anisotropies, one in plane and the other perpendicular to the film plane \cite{demag}. The latter can be incorporated into the interface perpendicular anisotropy field, forming an effective perpendicular anisotropy field. Inclusion of the former, in-plane demagnetization anisotropy in the form of a hard-axis anisotropy will only change the dispersion relation, making no difference to the understanding of the physics responsible for the formation of magnonic bands. Non-uniform dynamic demagnetization field \cite{Kruglyak07,Giesen07} can cause localization of SWs, and affect the SW modes in nanomagnets. Since we are interested in the influence on the magnonic dispersion relation of a periodic structure, we consider only the lowest lying mode with parabolic-like dispersion curves. Inclusion of higher modes will introduce more complex structures to the bands obtained with only the lowest mode, which is out of the scope of the current paper.} For the 1D geometry considered here (Fig. \ref{cell}), the gradient operator reduces to a differentiation on the $y$ variable, since the periodic structure of the MC is along the $y$-axis and we consider only magnetization variation along this direction.

In the ground state, which is derived from the static LLG equation, $\textbf {m}$ is uniformly magnetized along the easy ($z$) axis in region 1, $\textbf {m} = \hat {z}$. To obtain the SW eigenfunction, a single harmonic deviation from the uniform solution is considered, $\textbf {m} \propto \hat {z} + \mbox {\boldmath $\rho$} \exp (- i \omega t)$, where $\mbox {\boldmath $\rho$}$ is a vector in the film plane, thus perpendicular to the ground state magnetization direction $\hat {z}$. Substitute this form of $\textbf {m}$ back into the LLG equation and retain only the first order terms of $\mbox {\boldmath $\rho$}$, we can get the SW eigenequation
\begin{equation}
i \frac {\omega} {\omega_c} \mbox {\boldmath $\rho$} = \hat {z} \times (\delta^2 \ddot {\mbox {\boldmath $\rho$}} - \mbox {\boldmath $\rho$}),
\end{equation}
where the abbreviation $\ddot {\mbox {\boldmath $\rho$}} = d^2 \mbox {\boldmath $\rho$}/d y^2$ is employed to make the equation compact. Let $\phi = \rho_x + i \rho_y$, then the eigenequation for $\phi$ has a scalar form
\begin{equation}
- \frac {\omega} {\omega_c} \phi = (\delta^2 \ddot {\phi} - \phi).
\end{equation}
The SW dispersion relation $\omega/ \omega_c = 1 + \delta_1^2 k_1^2$ and the eigenfunction $\phi = \exp (i k_1 y)$ follows immediately. It is interesting to note that, except for the existence of an excitation gap, whose direct cause is the finite anisotropy field, the SW dispersion relation is identical to that of an electron travelling freely, whose effective mass is determined by the exchange constant and saturation magnetization only.

In region 2, the procedure to obtain the SW eigenfunction is essentially identical to that for region 1. The main difference is that, instead of a single domain state, there is a 2$\pi$ Bloch DW present. A $\pi$ DW can also be considered. We choose a 2$\pi$ DW simply because the corresponding unit cell is simpler. Since a detailed derivation of the magnetization profile for a confined DW and the corresponding SW eigenfunction is given in Ref. [\onlinecite{wangepl}], we will only describe the outline here. For brevity, in the following discussion of the DW magnetization profile and SW eigenfunction, the origin of the coordinate will be temporarily shifted to $y = d_2/4$. The ground state DW profile is described by a cosine Jacobian elliptic function \cite{A&S} $\cos \theta = -\mbox {sn} (y /\sqrt {\mu} \delta_2, \mu)$, where $\theta$ is the magnetization tilt angle in the $xz$ plane, measured from the $z$ axis, and $K(\mu)$ is the first kind complete elliptic integral. The modulus $\mu$ is determined by the length of region 2, $4 \delta_2 \sqrt {\mu} K(\mu) = d_2$. \textit{Due to the continuous rotation of the magnetization vector, the SW excitation is oscillating in a plane perpendicular to the local magnetization vector. Viewed in this perpendicular plane, the SW oscillation is identical to that of a single domain, which is discussed above, except that the oscillation now is elliptical. Hence, although the SW oscillation is vectorial globally, it is locally oscillating in a plane. The global 3D characteristics of the SW oscillation is restored if the rotation of the oscillating plane is considered. This feature of the SW excitation guarantees that we can use again a scalar function to describe SWs inside DWs. By rotating the $z$-axis to the local magnetization direction, and taking into account of the elliptical characteristics of the SW oscillation, the scalar SW eigenequation becomes}
\begin{equation}
- \left( 1 +  \frac {\mbox {dn}^2 (y_0, \mu)} {\mu} \right) \phi  =  \left( \ddot {\phi} \delta^2 - 2 \phi \, \mbox {sn}^2 \left( \frac {y} {\sqrt {\mu} \delta_2}, \mu \right) \right).
\label{dwequation}
\end{equation}
\textit{$y_0$ is an auxiliary constant related to $\omega$ through $\sqrt{\mu} \omega = \mbox{cn}(y_0, \mu) \mbox{dn}(y_0, \mu)$}, and sn is the sine Jacobian elliptic function \cite{A&S}. This equation has the form of a Schr\"{o}dinger equation with an elliptic potential, known mathematically as the Lam\'{e} equation \cite{Whittaker}. The same equation was obtained in the study of the excitation spectrum of the sine-Gordon equation \cite{Sutherland73}. The period of the potential in Eq. (\ref{dwequation}) is only half of the period of the magnetization distribution, which is $d_2$ for our case of a 2$\pi$ DW. This halving of the real space periodicity indicates a doubling of the $k$-space periodicity for SWs, which is 4$\pi/d_2$. This difference in periodicity of the SW and the static magnetization is caused by the insensitivity of SWs to DW chirality and polarity.
\begin{figure}\centering
\begin{minipage}[c]{0.6\linewidth}
\includegraphics[width=\linewidth]{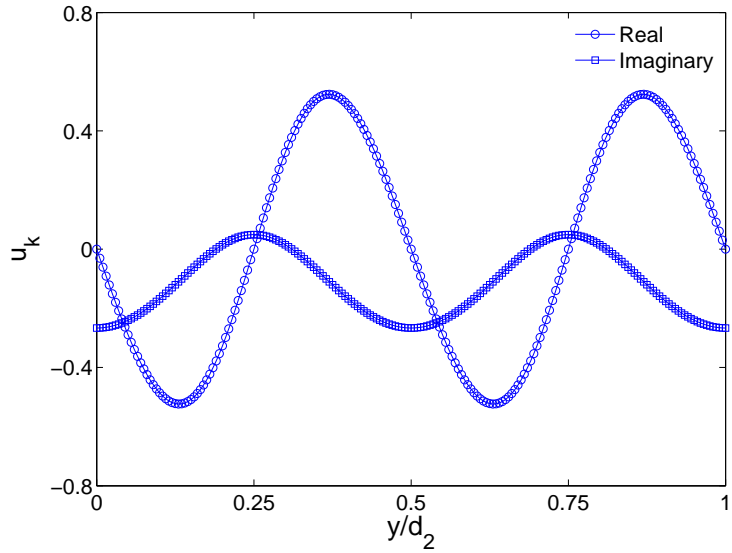}
\end{minipage}
\caption{Real and imaginary parts of the SW modulation function $u_k$ as a function of the spatial coordinate $y$ at 15 GHz. The total extension of the plotted region is $d_2 = 0.1 \mu$m. The DW width is $\delta_2$ = 20 nm.}
\label{wavefunction}
\end{figure}

The propagating solution of Eq. (\ref{dwequation}) is given by
\begin{equation}
\label{phi}
\phi = \frac{H(y/\sqrt {\mu} \delta_2 + y_0)}{\Theta(y /\sqrt {\mu} \delta_2)}e^{- y Z(y_0)/\sqrt {\mu} \delta_2},
\end{equation}
where $H$, $\Theta$ and $Z$ are Jacobi's eta, theta and zeta functions \cite{A&S}, respectively. For propagating SWs, $y_0$ is pure imaginary. To obtain the form of Eq. (\ref{phi}), periodic boundary condition was employed. From Eq. (\ref{phi}), it is obvious that the direct consequence of the introduction of magnetic DWs is to modulate the SW eigenfunction, which now has the form of a modulated plane wave. The origin of the modulation on the eigenfunction can be traced back to the periodic potential appeared in the eigenequation for the SW. Due to this periodic potential, the wavefunction should have the form of a Bloch wavefunction, $\phi (y) = u_k (y) e^{ i k_2 y}$, where $u_k$ is a periodic function with the period of the potential. To cast $\phi$ into this form, the crystal wave-vector has to be defined as $k_2 = i Z(y_0)/\sqrt {\mu} \delta_2 + 2 \pi/d_2$. The constant $2 \pi/d_2$ corresponds to the first Brillouin zone boundaries. Due to this displacement of the dispersion relation, there are two energy minima in the first Brillouin zone for SWs in the DW, and they are located at the zone boundaries. Interestingly, there is no band gaps opening in the whole Brillouin zone, although the potential function is periodic. The modulation function $u_k$ is given by $u_k = \exp(- 2 i \pi y/d_2) H(y/\sqrt {\mu} \delta_2 + y_0)/\Theta(y /\sqrt {\mu} \delta_2)$, which now satisfies $u_k (y + d_2/2) = u_k (y)$. The periodicity of $u_k$ can be seen from Fig. \ref{wavefunction}, where the real and imaginary parts of the modulation function $u_k$ at the frequency of 15 GHz are shown. It is interesting to note that, although $u_k$ is periodic with the same periodicity of the potential, its phase changes 2$\pi$ after a displacement of the period of the potential, which corresponds to a $\pi$ rotation of the magnetization vector. A 4$\pi$ phase is achieved when another period of the potential is displaced, and the magnetization rotation is correspondingly 2$\pi$. From this observation, the phase shift of $u_k$ and the magnetization rotation of the DW is proportional to each other. The phase of $u_k$, $\varphi$, is a Berry phase \cite{Bruno04, Guslienko10}, in contrast to the dynamical phase related to the crystal momentum. The magnetization rotation divided by 2$\pi$ is defined as the winding number \cite{braun} $n_w$ in the configuration space. Therefore, the Berry phase of SWs in the DW is proportional to the winding number, which is a topological quantity. The relation between $\varphi$ and the winding number, $\varphi = 4 n_w \pi$, confirms that the SW Berry phase inherits the topological character of the underlying DW.

The potential in the Schr\"{o}dinger equation, Eq. (\ref{dwequation}), and hence the whole Hamiltonian, are invariant under spatial inversion, which implies that the eigenfunction can be chosen to have definite parity. The function $\phi$, however, does not have a definite parity. Under the operation of spatial inversion, $\phi$ is transformed to $-\phi(-y_0) = - \phi^*$, with $\phi^*$ the complex conjugate of $\phi$. Eigenfunctions with definite parity can be constructed from $\phi$ and $\phi^*$. Due to the CPT theorem \cite{CPT}, consecutive action of time reversal and spatial inversion will leave $\phi$ intact, given that magnons are charge neutral particles. Hence, the Hamiltonian for SWs in the DW is invariant under time-reversal operation, and $\phi$ and $-\phi^*$ form time-reversal and space-inversion pairs. Due to the time reversal symmetry, or equivalently, the inversion symmetry, the Berry phase of SWs with opposite $k$ will have opposite signs, signifying the different topological charge associated with positive or negative $k$. As far as time reversal symmetry is conserved, the back scattering of magnons at the Brillouin boundaries will be prohibited, thus forbidding the corresponding band gap opening. This is consistent with the calculated SW energy spectrum. \textit{At this point, it should be emphasized that the dependence of the Berry phase on the crystal momentum is the consequence of the conservation of the time reversal symmetry, in contrast to the non-reciprocity of SW propagation \cite{Verba,Lisenkov}, which requires the violation of the time reversal symmetry.}

The SW Berry phase $\varphi$ can be expressed as the line integral of a field $\beta(y)$, $\varphi =  - \int dy \beta$. From $\beta$, a vector potential is derivable. The SW eigenfunction can be rewritten as $\phi =  \exp { i (\varphi + k_2 y) }$. With the definition of $\beta$, it can be easily seen that $\phi$ satisfies a simple equation, $(\partial/\partial y + i \beta) \phi = i k_2 \phi$, which enables us to define a covariant derivative operator \cite{Greiner}, $D = \partial/\partial y + i \beta$. The explicit form of the vector potential is $\beta = 2 \pi/d_2 + i(Z(y/\sqrt{\mu} \delta_2 + y_0) - Z(y/\sqrt{\mu} \delta_2) + \mbox{cs} (y/\sqrt{\mu} \delta_2 + y_0) \mbox{dn} (y/\sqrt{\mu} \delta_2 + y_0))/\delta_2 \sqrt{\mu}$. Potential-like field $\beta$ is generally not a real-valued field. Only when the magnetization is pointing along the easy axis, $\beta$ is a pure real number. $\phi$ is the eigenfunction of the covariant derivative operator with the eigenvalue $ik$, although $\phi$ is not an eigenfunction of the real derivative operator, $\partial/ \partial y$. It is interesting to note that, instead of the real space, if we consider the configuration space of the magnetization $(m_x, m_z)$, the Berry phase can be expressed as the contour integral of a vector potential-like field in the configuration space, $\varphi =  - \oint d\theta \beta_\theta$. The topological character of the Berry phase is easier to observe in the 2D $(m_x, m_z)$ configuration space. As we will see in the next section, the difference between the crystal momentum and the vector potential-like field $\beta$ can be observed as the magnonic band structure reconfiguration induced by the presence of the DW in the unit cell.

\section{Band structure of domain-wall magnonic crystals}
\label{band}
With those SW eigenmodes in regions 1 and 2 known, we are now ready to discuss the band structure of the whole MC. We use the same method employed previously for the discussion of the SW spectrum \cite{Albuquerque86, Albuquerque92,Barnas92,Vasseur96,Wang15} in periodic magnetic structures. Micromagnetic simulations can also be used to deal with the same problem, similar to the treatment of magnonic band structure in a width-modulated MC \cite{Lee09}. While micromagnetic simulations can be more realistic, including non-uniform dipolar field and pinning effects, the analytic method employed here can provide more insight into the physical mechanism responsible for the magnonic band formation. For the MC considered here, Bloch theorem requires that the eigenfunctions in both regions have the form
\begin{equation}
\psi_i = e^ {i k y} u_i(y), i = 1 \mbox { or } 2,
\end{equation}
where $k$ is the Bloch wave vector, confined to the first Brillouin zone $[-\pi/ d, \pi/ d]$. $d$ is the MC's period and $u_i$ are arbitrary functions with the same period of the MC, $u_i(y + n d ) = u_i (y)$, where $n$ is an integer. $\psi_i$ here can be chosen as a linear combination of the SW eigenfunctions, $\psi_1 = a_1 \exp (i k_1 y) + b_1 \exp (-i k_1 y)$ in region 1 and $\psi_2 = a_2 \phi + b_2 \phi^*$ in region 2. The periodicity of $u_i$ will be guaranteed by a suitable choice of the coefficients $a_i$ and $b_i$. Since the eigenequation for $\psi_i$ is identical in form to a 1D Schr\"{o}dinger equation, coefficients $a_i$ and $b_i$ can be obtained by imposing the periodic boundary conditions for a Schr\"{o}dinger equation. Specifically, this means that $u_1 (0) = u_2 (0)$ and $\dot{u}_1 (0) = \dot {u}_2 (0)$ at the origin (which is the central interface). Periodicity is guaranteed by the boundary conditions at the outer interfaces, $u_1 (- d_1) = u_2 (d_2)$ and $\dot {u}_1 ( - d_1) = \dot {u}_2 ( d_2)$. \textit{As discussed in Sec. \ref{sws}, the SW oscillation inside a DW is actually 3D, and the scalar eigenfunction represents the oscillation in a plane perpendicular to the local magnetization vector. The boundary conditions employed here for the scalar functions in both regions 1 and 2 ensure that the spin current is conserved inside the unit cell. The continuity of the magnetization across the boundaries provides the applicability of the boundary conditions given above, which require that the SW oscillation is in the same plane}. Correspondingly, the secular equation gives an implicit equation to determine the band structure
\begin{equation}
\label{band-wall}
\cos k d = \cos k_1 d_1 \cos (k_2) d_2 - \frac {k_1^2 + q^2} {2 k_1 q} \sin k_1 d_1 \sin k_2 d_2.
\end{equation}
As derived in Sec. \ref{sws}, $k_1$ is related to the frequency through $\omega/ \omega_{c, 1} = 1 + \delta_1^2 k_1^2$, which is the well-known dispersion relation for SWs in a single domain state. In the DW, however, the SW dispersion relation is not that simple any more, $\omega/ \omega_{c, 2} = \mbox {dn} (\alpha, \mu_1)/ \sqrt {\mu} \mbox {cn}^2 (\alpha, \mu_1)$. $\alpha$ is a real parameter related to $y_0$ through $y_0 = i \alpha$, and $\mu_1$ is the complementary modulus, $\mu + \mu_1 = 1$. The crystal momentum in region 2 is given by $k_2 \delta_2 \sqrt {\mu}= Z ( \alpha, \mu_1) + \alpha \pi / 2 K(\mu) K'(\mu) - \mbox {sc} ( \alpha, \mu_1) \mbox {dn} ( \alpha, \mu_1)$. The constant $2\pi/d_2$ is omitted because it makes no contribution to Eq. (\ref{band-wall}). $K'(\mu) = K(\mu_1)$ is the complementary first kind complete elliptic integral and $q \delta_2 \sqrt {\mu} = - \mbox {sc} ( \alpha, \mu_1) \mbox {dn} ( \alpha, \mu_1)$. For comparison, the corresponding band structure for the MC structure without the 2$\pi$ DW is determined by the following equation
\begin{equation}
\label{band-domain}
\cos kd = \cos k_1 d_1 \cos k_2 d_2 - \frac {k_1^2 + k_2^2} {2 k_1 k_2} \sin k_1 d_1 \sin k_2 d_2,
\end{equation}
which is similar to the equation derived in the classical Kronig-Penney model \cite{kpmodel}. This similarity is self-evident: In both cases, we use the same Schr\"{o}dinger equation and Bloch theorem. Wave-vectors $k_i$ are related to the frequency through $\omega/ \omega_c ^i = 1 + \delta_i^2 k_i^2$ for this domain MC. If the two regions have identical material parameters, then the two wave vectors are equal to each other, $k_1 = k_2$, and the crystal wave vector $k$ reduces to the real wave vector, which means that there is no band gaps for a continuous film.

Compared to Eq. (\ref{band-domain}), Eq. (\ref{band-wall}) shows that the main effect of the DW is to modify the momentum factor appearing in the implicit equation for the determination of the band structure. This modification can be understood on the fact that, due to the presence of the DW, the SW's crystal momentum and linear momentum are not the same quantity anymore. The crystal momentum of SWs in the DW is $k_2$, which corresponds to the eigenvalue of the covariant derivative operator. When the SW eigenfunction is displaced by $d_2$, its phase change is $k_2 d_2$, in analogy to the real linear momentum in continuous space. A similar conclusion was reached using micromagnetic simulation \cite{Hertel04}. The linear momentum operator is proportional to the gradient $- i \nabla$, which reduces to $- i \partial/ \partial y$ in our 1D geometry. If there is no presence of the DW, a differentiation on the wavefunction gives a constant linear momentum. In the presence of the DW, the same differentiation on the corresponding SW wavefunction gives again the linear momentum. But the linear momentum is not a constant anymore. The linear momentum at the outer boundaries is given by $q = k_2 - \beta$, as compared to the crystal momentum $k_2$. This linear momentum will enter the momentum factor in the secular equation for the band structure. The disparity between the linear and crystal momenta derives from the difference between the ordinary and covalent differentiation operators \cite{Greiner} acting on the SW wavefunction, as discussed in Sec. \ref{sws}.
\begin{figure}\centering
\begin{minipage}[c]{0.6\linewidth}
\includegraphics[width=\linewidth]{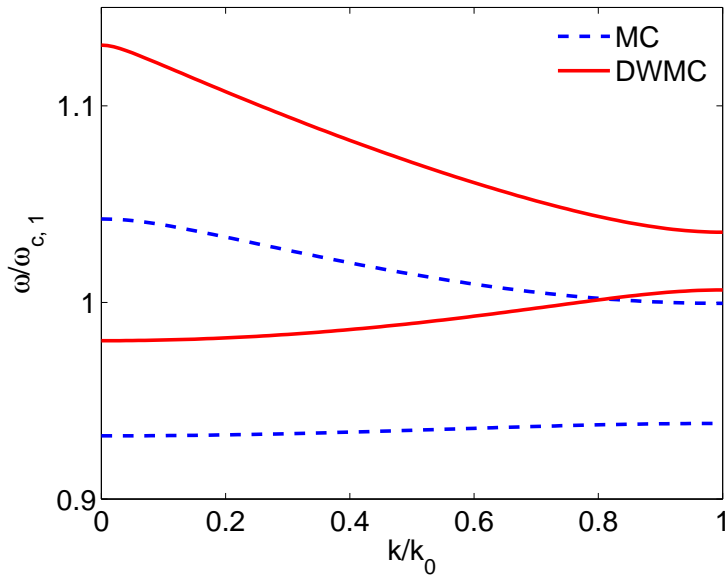}
\end{minipage}
\caption{The first two magnonic bands in the positive half of the first Brillouin zone with symmetric unit cell. The solid (red) and dashed (blue) lines are for the DWMC and a reference single domain MC, respectively. The unit cell size is $d = d_1 + d_2 = 0.4\; \mu$m, with equal lengths for the two regions $d_1 = d_2$. $k_0$ is the Brillouin zone boundary, $k_0 = \pi/d$. Higher energy bands have smaller band gaps, and are not shown. The bands of the same DWMC but with identical magnetic parameters in the two regions are shown as dash-dotted (red) lines, demonstrating the persistence of finite band gap caused by the presence of the DW in the unit cell.}
\label{sym-band}
\end{figure}

To get numerical values, we need to specify the magnetic parameters. The magnetic parameters in the two regions should be different to get sizable energy gaps, which can be realized by ion implantation \cite{Franken12,Buckingham12} or using artificial superlattices \cite{Wang09,Qiu14}. However, it should be noted that, in contrast to conventional 1D MCs, DWMCs have energy gaps even if regions 1 and 2 have the same set of magnetic parameters (cf. Fig. 1, dash-dotted lines), due to the different SW dispersion relation caused by the DW. The disappearance of magnonic band gaps in conventional MCs is due to the restoration of the translation invariance, if the two regions have identical magnetic parameters. In the presence of the DW, the translation invariance is not restored even if the two regions are identical, which explains why there are still finite band gaps in this case. It should be noted that this explanation is based on the unit cell structure shown in Fig. \ref{cell}, where the DW is confined to region 2. In the case of identical magnetic parameters for the two regions, the DW will expand to occupy the whole unit cell, and the corresponding band gaps will vanish if this effect is considered. This effect is ignored in all of our calculations, and we stick strictly to the unit cell given in Fig. \ref{cell}. As only $\omega_c$ and $\delta$ enter Eq. (\ref{band-wall}), we do not need to specify all three parameters, $A$, $K$ and $M_s$. For region 1, we choose $\omega_{c, 1} = 10$ GHz and $\delta_1 = 20$ nm. We assume that, either due to ion implantation or material combination, in the second material, the anisotropy constant is reduced by a value of 10\%. Other parameters remain the same in region 2. With those parameters specified, the band structure can be computed from Eq. (\ref{band-wall}) directly. An example band structure is shown in Fig. \ref{sym-band}. As can be expected, the appearance of energy bands and band gaps is obvious. Only the first two bands with a significant band gap between them are shown. All the other bands with higher band index have negligible band gaps.

From Fig. \ref{sym-band}, we can see that the inclusion of the DW in the unit cell has significant effects. To facilitate a direct comparison, we use the same set of magnetic parameters to get the band structure for a conventional MC with the same unit cell, which is also given in Fig. \ref{sym-band}. Without the DW, the first band is very flat, meaning the group velocity there is very small. If the DW is present, the group velocity is increased. For higher energy bands, this effect is relatively less important. This transition from slow to fast propagation of SWs in the first band is actually caused by the increased cut-off frequency in the DW. For a DW characterized by the modulus $\mu$, the cut-off frequency is $\omega_0/ \sqrt {\mu}$, scaled up by a factor of $1/\sqrt {\mu}$, compared to the single domain case. In the first band, the SW is evanescent in region 2 for the domain MC, while it is propagating for the case of a DWMC. In addition, band gap and gap position can both be tuned by the sole presence of the DW, as shown in Fig. \ref{sym-band}. This signifies the main advantage of employing magnetization textures, whose representative is a DW, in the unit cell of a MC: Application of external field, either electric or magnetic, can tune the band structure. For the case considered in Fig. \ref{sym-band}, the application of a magnetic field parallel to the $z$ direction can annihilate the DW, hence collapsing the band structure to that of a domain MC. After this transition, some forbidden states in the band gap are allowed to propagate, realising reconfigurable control over SWs' propogation. Application of a external magnetic field parallel to the $-z$ direction will modify the magnetization profile, thus affecting the SW characteristics and band structure, which will be investigated in the future.
\begin{figure}\centering
\begin{minipage}[c]{0.45\linewidth}
\includegraphics[width=\linewidth]{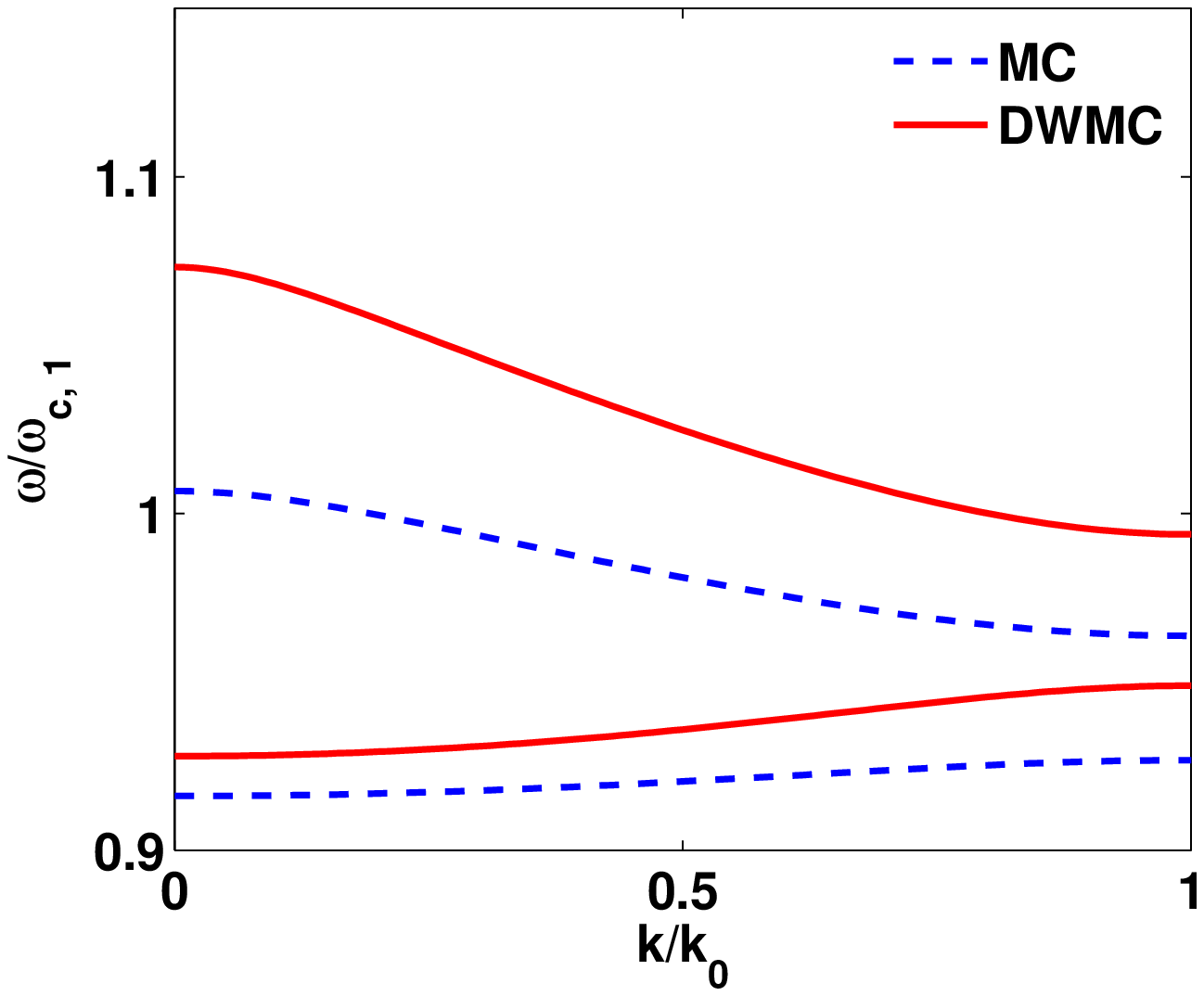}
\end{minipage}
\hfill
\begin{minipage}[c]{0.45\linewidth}
\includegraphics[width=\linewidth]{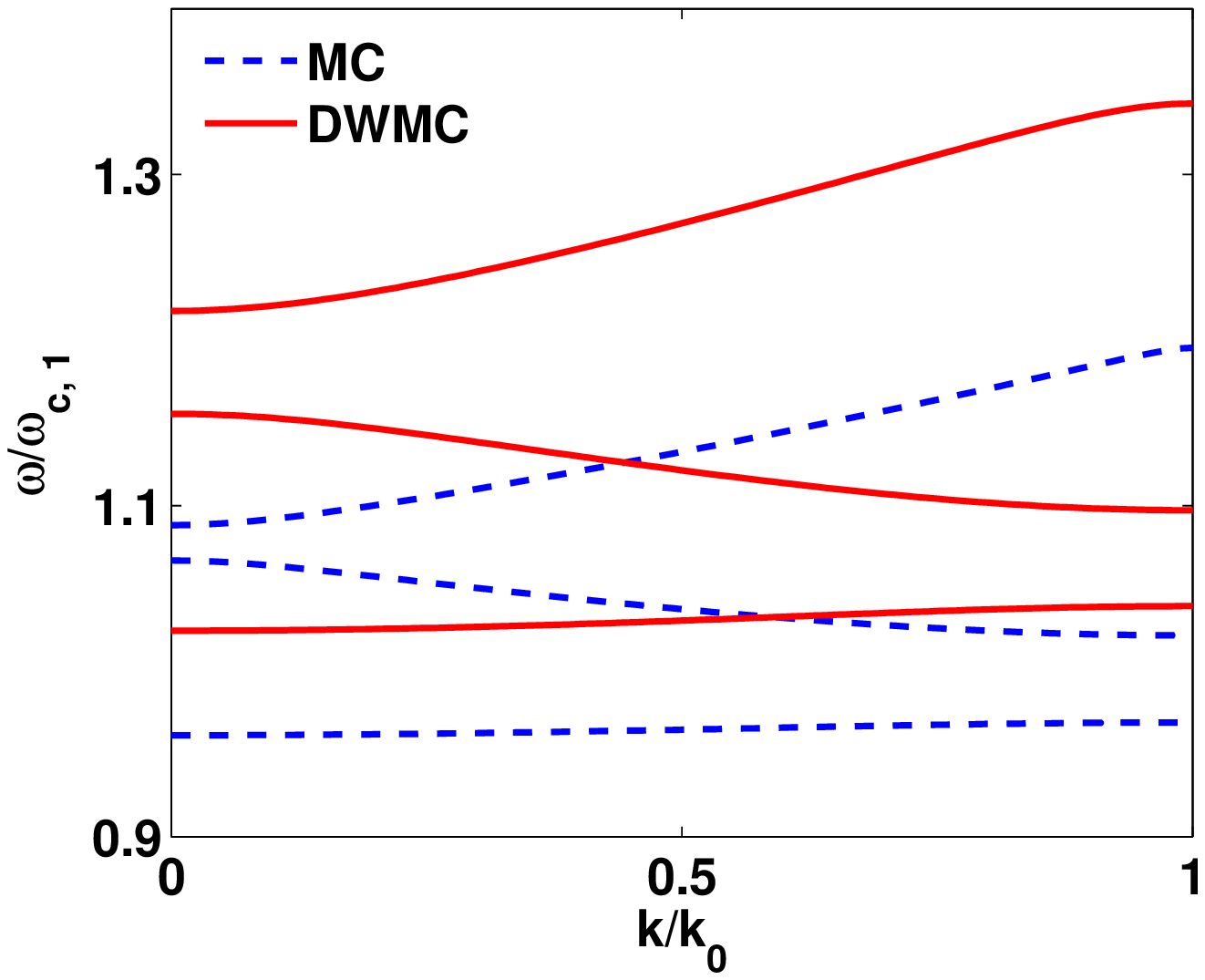}
\end{minipage}
\caption{Magnonic energy bands in the positive half of the first Brillouin zone with asymmetric unit cell. The solid (red) and dashed (blue) lines are for the DWMCs and corresponding reference single domain MCs, respectively. The cell size is $d = d_1 + d_2 = 0.4 \;\mu$m, with $d_2 = 3 d_1$ (left) and $d_1 = 3 d_2$ (right). $k_0$ is the Brillouin zone boundary, $k_0 = \pi/d$. Note the different scales for the frequency axes. The band gaps of higher energy bands are smaller, hence not shown.}
\label{asym-band}
\end{figure}

To further illustrate the versatility of the DWMC, we consider unit cells with unequal widths for the comprising pieces, but fixed total cell size. In Fig. \ref{asym-band}, the band structure of two asymmetric unit cells is given. It follows immediately that an expansion in the size of region 2, and hence a shrink in the size of region 1, has only quantitative significance. In contrast, a shrink in the size of region 2 changes the band structure qualitatively: there are only two bands with sizable band gaps for $d_2 =$ 0.3 $\mu$m, but that number increases to three with $d_2$ decreasing to 0.1 $\mu$m. This variation with $d_2$ is easily understood from the SW eigenequation. In a uniform domain, it is a Schr\"{o}dinger equation with a constant potential. In the presence of the DW, the potential varies with position. A decrease in $d_2$ tightens the variation of the potential. As is well known from electronic band theory \cite{kpmodel}, the energy gap is related to the Fourier components of the potential. Hence, a change in the potential will definitely affect the band structure. This observation lends further support to the claim that the main advantage of DWMCs is to offer, besides the conventional modulation of magnetic parameters, an additional degree of tunability due to the adjustable magnetization profile in the unit cell.

\begin{figure}\centering
\begin{minipage}[c]{0.4\linewidth}
\includegraphics[width=\linewidth]{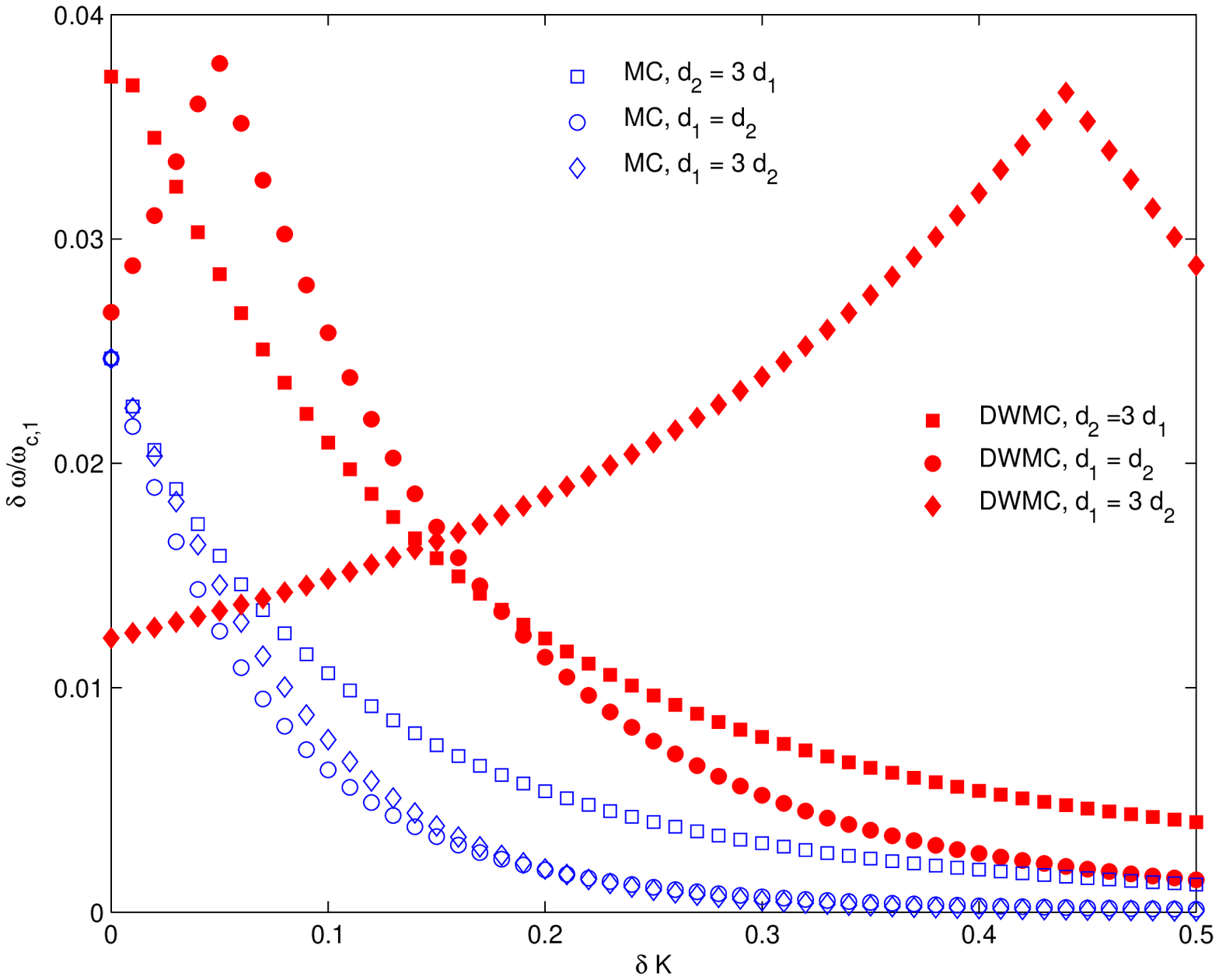}
\end{minipage}
\hfill
\begin{minipage}[c]{0.4\linewidth}
\includegraphics[width=\linewidth]{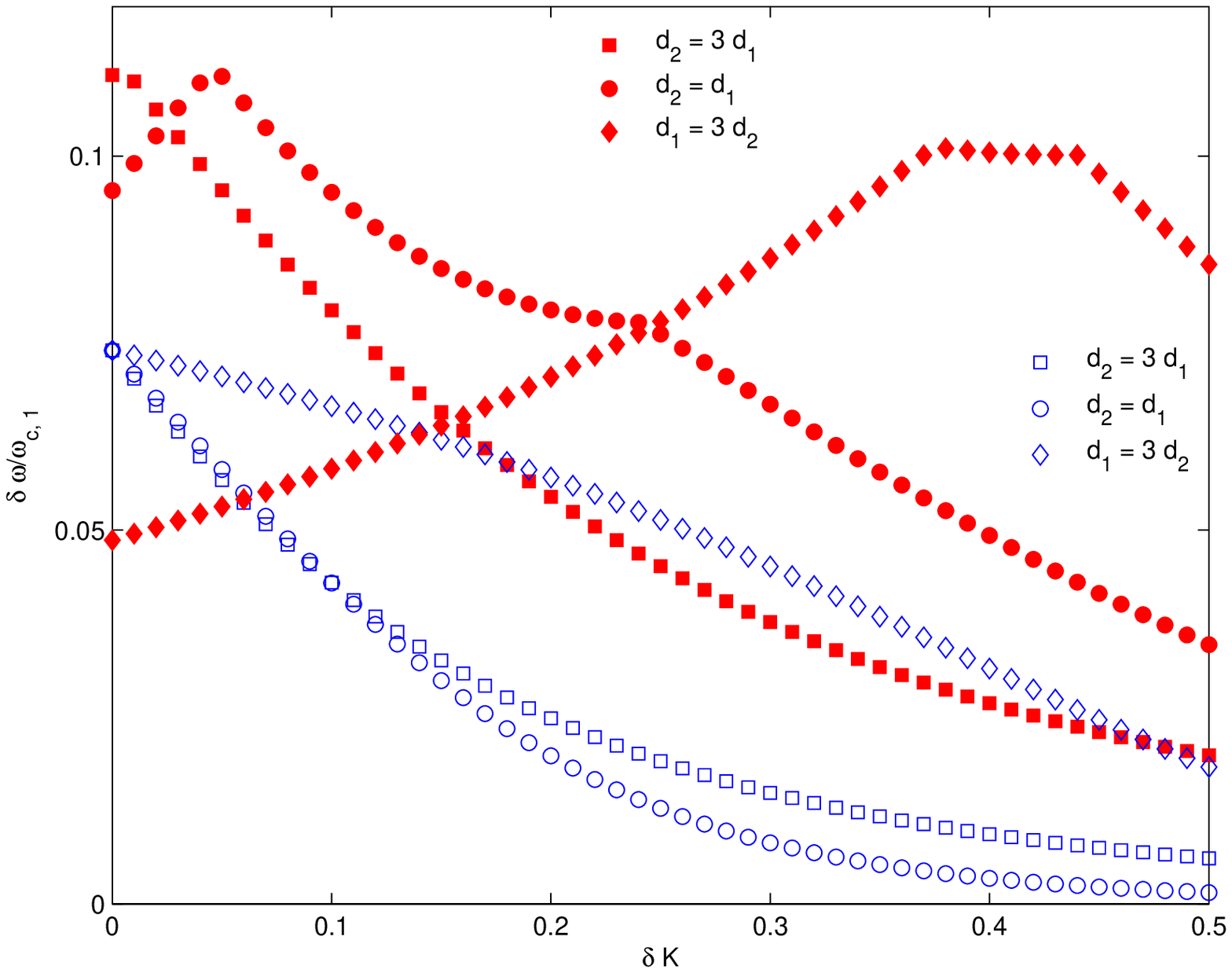}
\end{minipage}
\caption{Magnonic bandwidths of the first (left) and second (right) bands as a function of the reduction in the perpendicular magnetic anisotropy, $\delta K$, for fixed length of the unit cell $d = 0.4 \mu$m, but different ratios between the two regions inside the unit cell: $d_2 = 3 d_1$ (square), $d_2 = d_1$ (circle) and $d_1 = 3 d_2$ (diamond). Filled (red) and open (blue) symbols denote results for the DWMCs and reference single domain MCs, respectively.}
\label{bandwidth-k}
\end{figure}

\section{band gaps and band widths}
\label{gap}
For applications, the band width and band gap are important parameters. Fig. \ref{bandwidth-k} shows the variation of the band width for the first and second bands, as a function of the change in anisotropy, $\delta K = 1 - K_2/K_1$. For domain MC, it is obvious that the band width of both the first and the second bands decreases with $\delta K$. At the value $\delta K = 0$, the band width is given by the relation $\delta \omega = (2n - 1) \pi^2/d^2$, with $ n$ the band index, derived from the dispersion relation. Since the unit cell length $d$ is the same in Fig. \ref{bandwidth-k}, the three curves should converge to the same point at $\delta K = 0$. At this point, the band gap is zero, since there is no modulation of anisotropy and the translational invariance is restored. In contrast, the band width for DWMC is not a monotonous function of $\delta K$. The effect of the nucleation of the DW can be observed conspicuously when the DW is more tightly confined: the band width increases first, then decreases with the increase of $\delta K$. For the band width of the second band, its behavior is even more complex. There is a transition region, connecting the increasing and decreasing parts. If the length of region 2 is large, the effect of the DW is not that significant, and the band width decreases with $\delta K$, similar to the case of domain MCs. The band width of a band is determined by both the SW dispersion relations and the modulation in anisotropy in the unit cell. There is no simple rules to determine the behavior of the band widths before carrying out a numerical calculation. The qualitative behavior, however, is predictable. For example, when $\delta K$ is large, both the band widths of the first and second bands become small, since both bands are derived from evanescent waves in region 1. The resulting bands are narrow and have little dispersion, inherent of the localized characteristic of evanescent waves.
\begin{figure}\centering
\begin{minipage}[c]{0.4\linewidth}
\includegraphics[width=\linewidth]{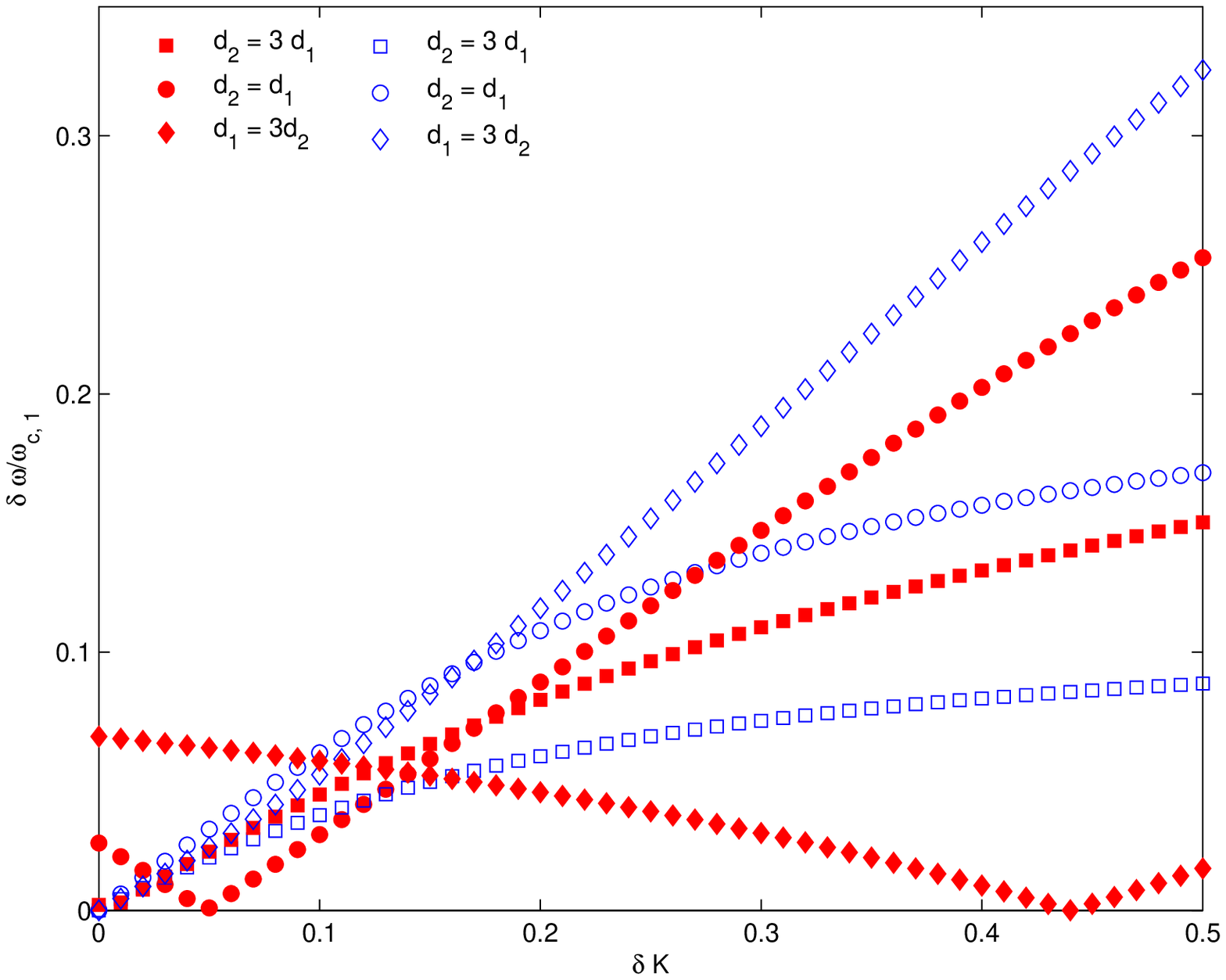}
\end{minipage}
\hfill
\begin{minipage}[c]{0.4\linewidth}
\includegraphics[width=\linewidth]{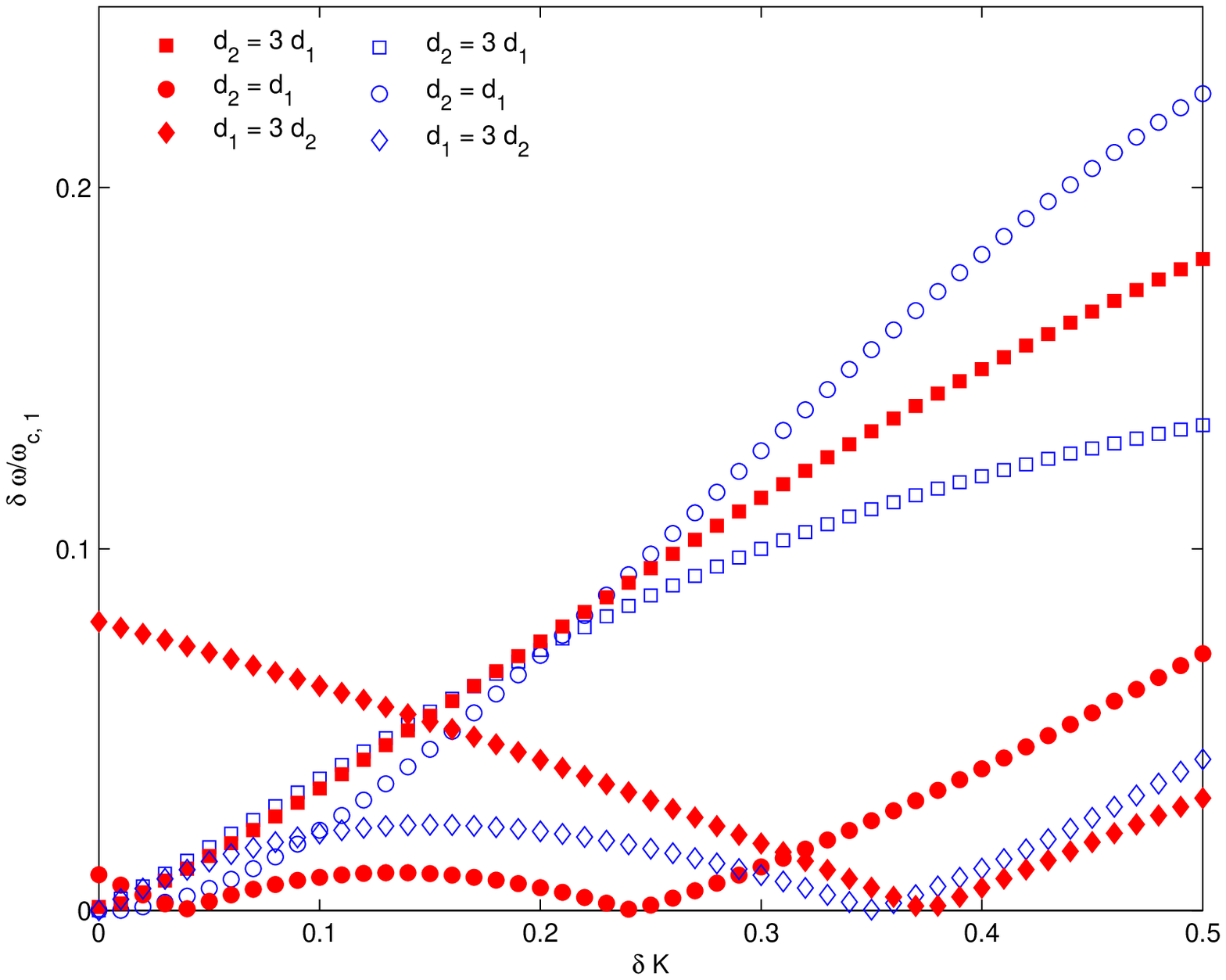}
\end{minipage}
\caption{Magnonic band gaps of the first (left) and second (right) bands as a function of the reduction in the perpendicular magnetic anisotropy, $\delta K$, for fixed length of the unit cell $d = 0.4 \mu$m, but different ratios between the two regions inside the unit cell: $d_2 = 3 d_1$ (square), $d_2 = d_1$ (circle) and $d_1 = 3 d_2$ (diamond). Filled (red) and open (blue) symbols denote results for the DWMCs and reference single domain MCs, respectively.}
\label{bandgap-k}
\end{figure}

As a function of the modulation in anisotropy, the band gaps for the first and second bands are shown in Fig. \ref{bandgap-k}. The band gaps of MC increases monotonously with $\delta K$, with inflection points in the plotted range for $\delta K$. In the presence of the DW in region 2, the band gaps are no longer zero when the modulation in $K$ is absent, which is obviously derived from the violation of the translational invariance, due to the mere presence of the DW. The finite band gap with no change of material parameters of DWMCs is in stark contrast to conventional domain MCs. For small $d_2$, there is a striking feature emerging, which is common to both types of MCs: the band gap closes once or twice in the considered range of modulation in anisotropy, according to the geometrical structure of the unit cell, indicating the disappearance of band gaps correspondingly. The sufficient and necessary condition to obtain a zero band gap is easily obtained from Eqs. (\ref{band-wall}) and (\ref{band-domain}), which is
\begin{equation}
k_1 d_1 = n \pi, k_2 d_2 = m \pi
\end{equation}
for both types of MCs. The same integer $m = n$ gives a zero band gap at the zone center, and the zero band gap will appear at the Brilloun zone boundaries if $m \neq n$. The condition can be interpreted as the condition for the wavefunction to interfere constructively, after propagating forward and reflected backward in each region, regions 1 and 2. Or equivalently, the length of each region in the unit cell is an integer multiple of half of the wave length, which is similar in form to the condition for the formation of confined, discrete energy levels in a potential well. If the condition is fulfilled, the reflectivity of each individual part of the unit cell is zero, consistent with the constructive interference condition.
\begin{figure}\centering
\begin{minipage}[c]{0.4\linewidth}
\includegraphics[width=\linewidth]{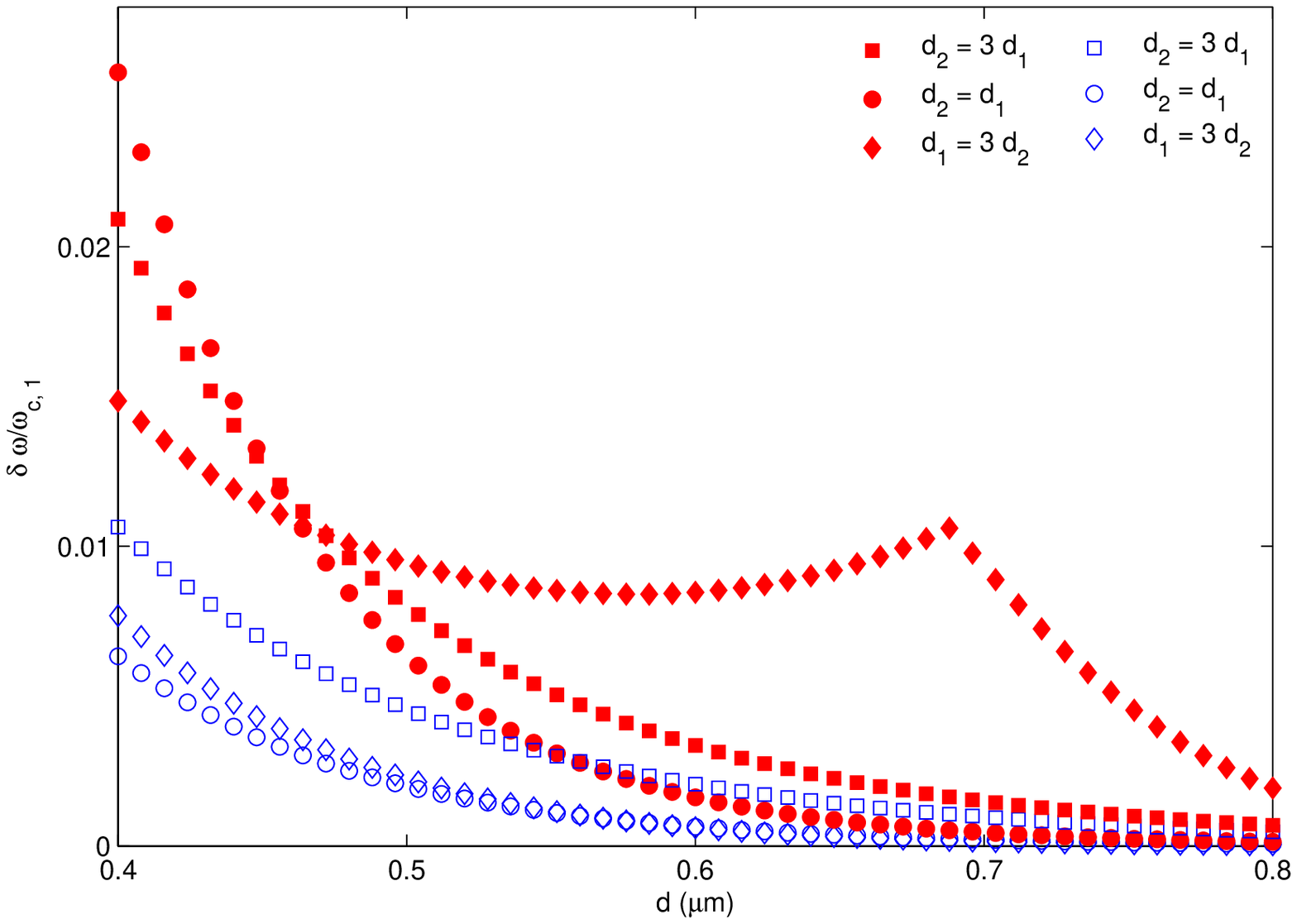}
\end{minipage}
\hfill
\begin{minipage}[c]{0.4\linewidth}
\includegraphics[width=\linewidth]{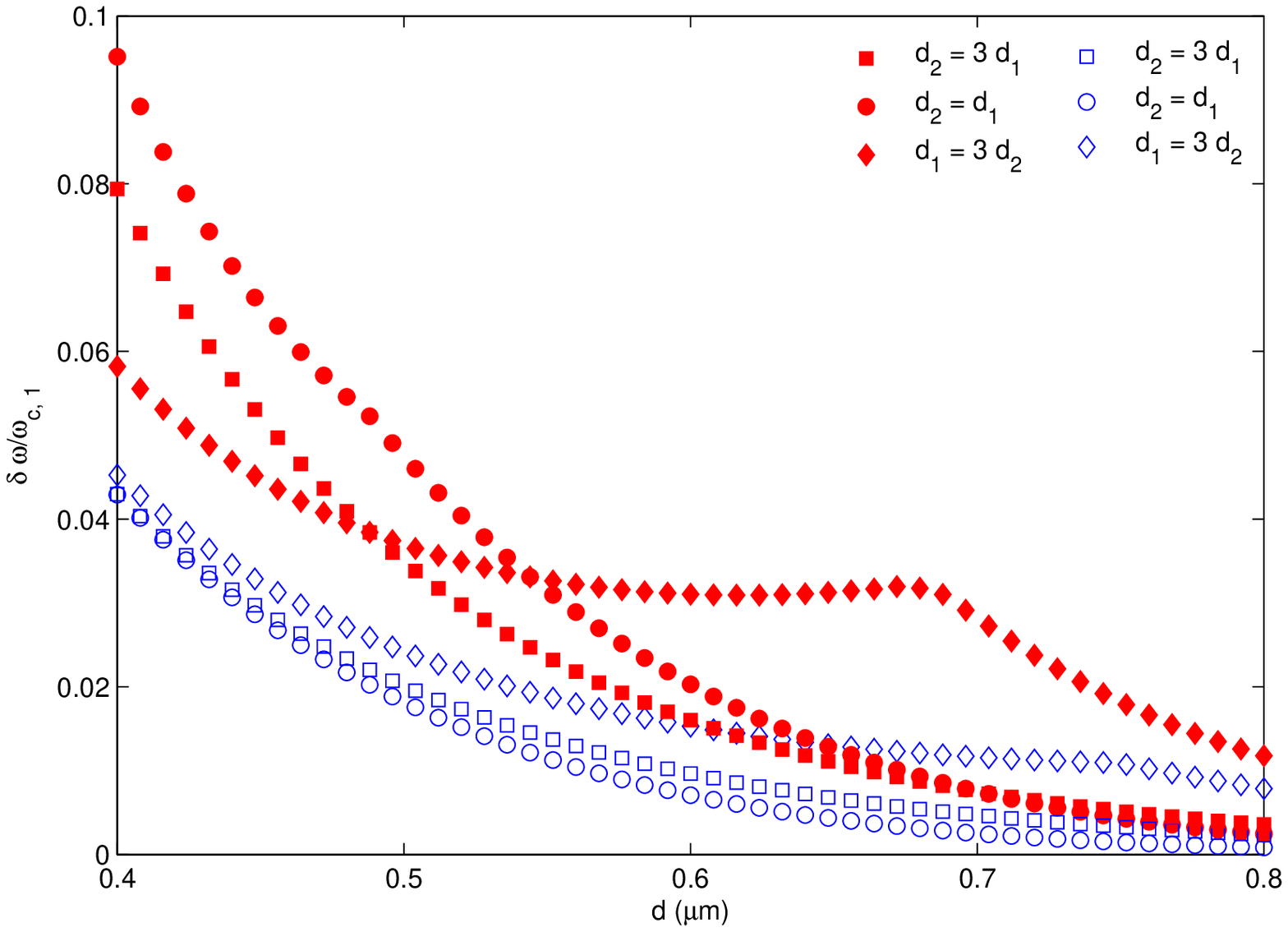}
\end{minipage}
\caption{Magnonic bandwidths of the first (left) and second (right) bands as a function of the length of the unit cell $d$, with three different ratios between the two regions inside the unit cell: $d_2 = 3 d_1$ (square), $d_2 = d_1$ (circle) and $d_1 = 3 d_2$ (diamond). Filled (red) and open (blue) symbols denote results for the DWMC and reference single domain MCs, respectively.}
\label{bandwidth-d}
\end{figure}

\begin{figure}\centering
\begin{minipage}[c]{0.4\linewidth}
\includegraphics[width=\linewidth]{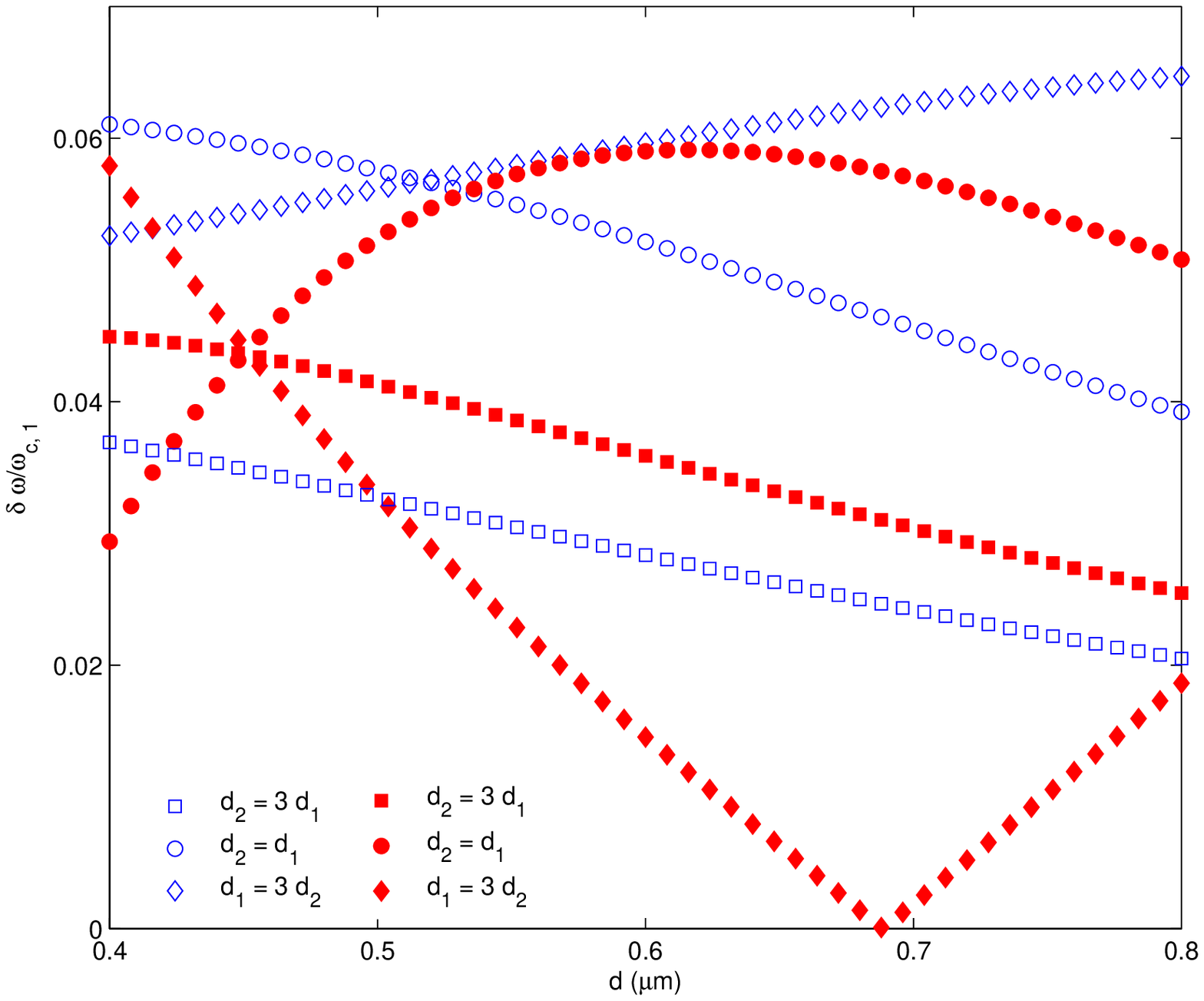}
\end{minipage}
\hfill
\begin{minipage}[c]{0.4\linewidth}
\includegraphics[width=\linewidth]{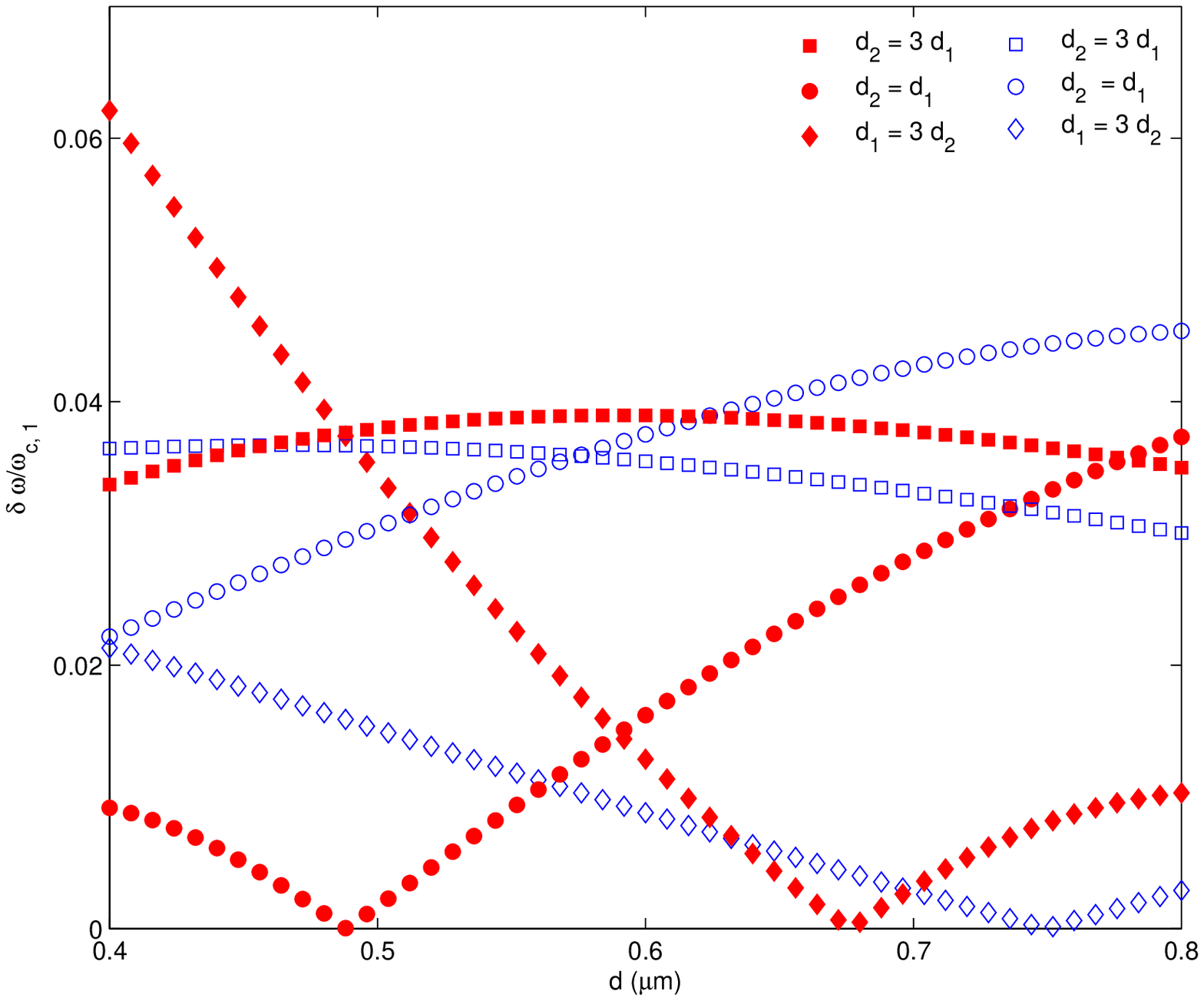}
\end{minipage}
\caption{Magnonic bandgaps of the first (left) and second (right) bands as a function of the length of the unit cell $d$, with three different ratios between the two regions inside the unit cell: $d_2 = 3 d_1$ (square), $d_2 = d_1$ (circle) and $d_1 = 3 d_2$ (diamond). Filled (red) and open (blue) symbols denote results for the DWMCs and reference single domain MCs, respectively.}
\label{bandgap-d}
\end{figure}

Similar behavior of band width (Fig. \ref{bandwidth-d}) and band gap (Fig. \ref{bandgap-d}) can be observed as the unit cell size $d$ is changed, while holding the ratio of the lengths of the two comprising parts in the unit cell constant. In contrast to the variation in anisotropy, a modification in $d$ will not change the dispersion relation for each part in the unit cell of a domain MC. The only effect of the variation in $d$ is to change the periodicity of the MC, and correspondingly the Brillouin zone boundaries. Therefore, the Fourier components involved in the determination of the band structure will change correspondingly, making the observed variations. For the DWMC, in addition to the modified periodicity, the dispersion relation of the DW part in the unit cell is affected by a change in $d$. This can be easily seen from the SW eigenequation Eq. (\ref{dwequation}), since the modulus $\mu$ is determined by $d_2$, which is proportional to $d$, and the dispersion relation is determined by $\mu$.

\section{development of the band strcture}
\label{evol}
\begin{figure}\centering
\begin{minipage}[c]{0.4\linewidth}
\includegraphics[width=\linewidth]{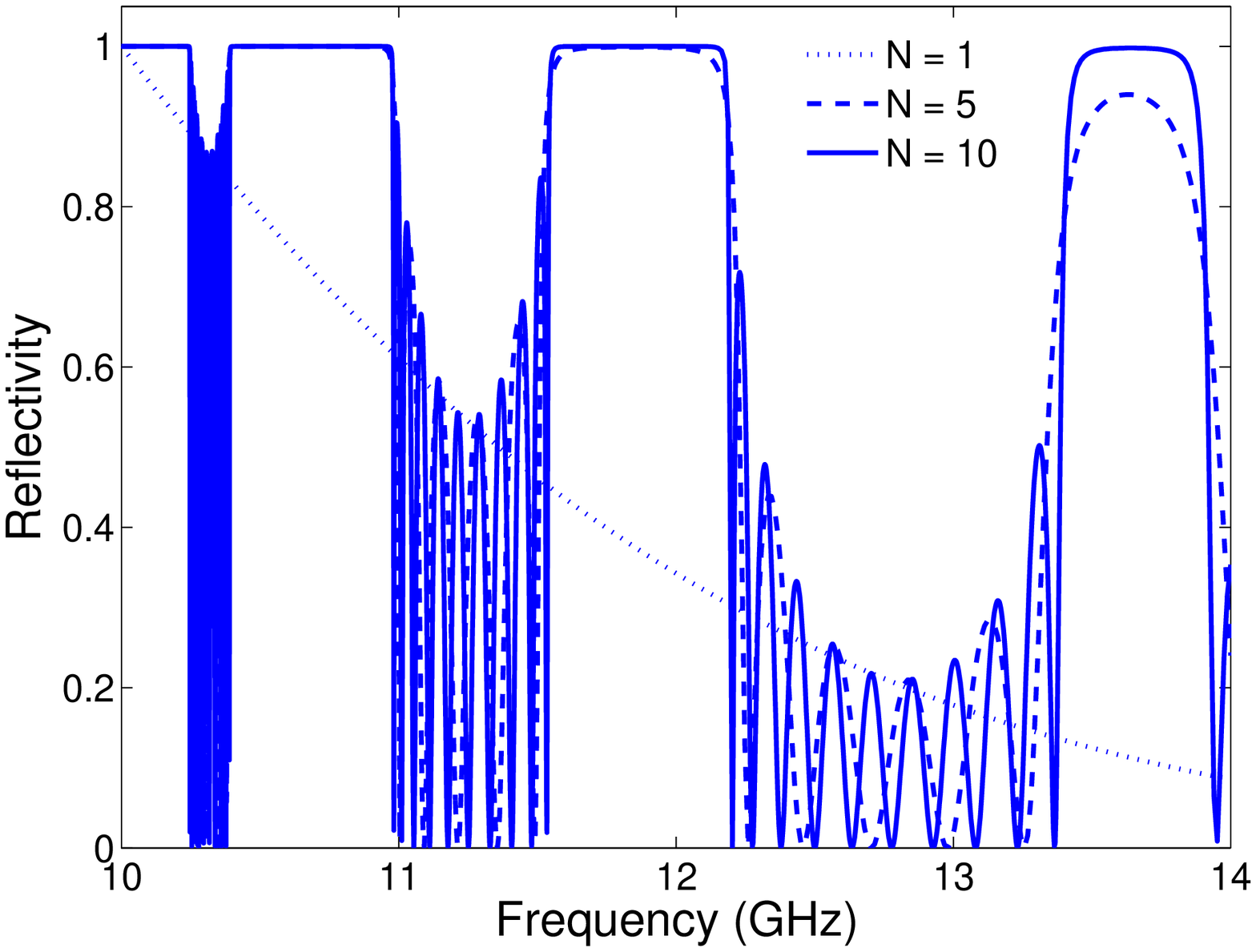}
\end{minipage}
\hfill
\begin{minipage}[c]{0.4\linewidth}
\includegraphics[width=\linewidth]{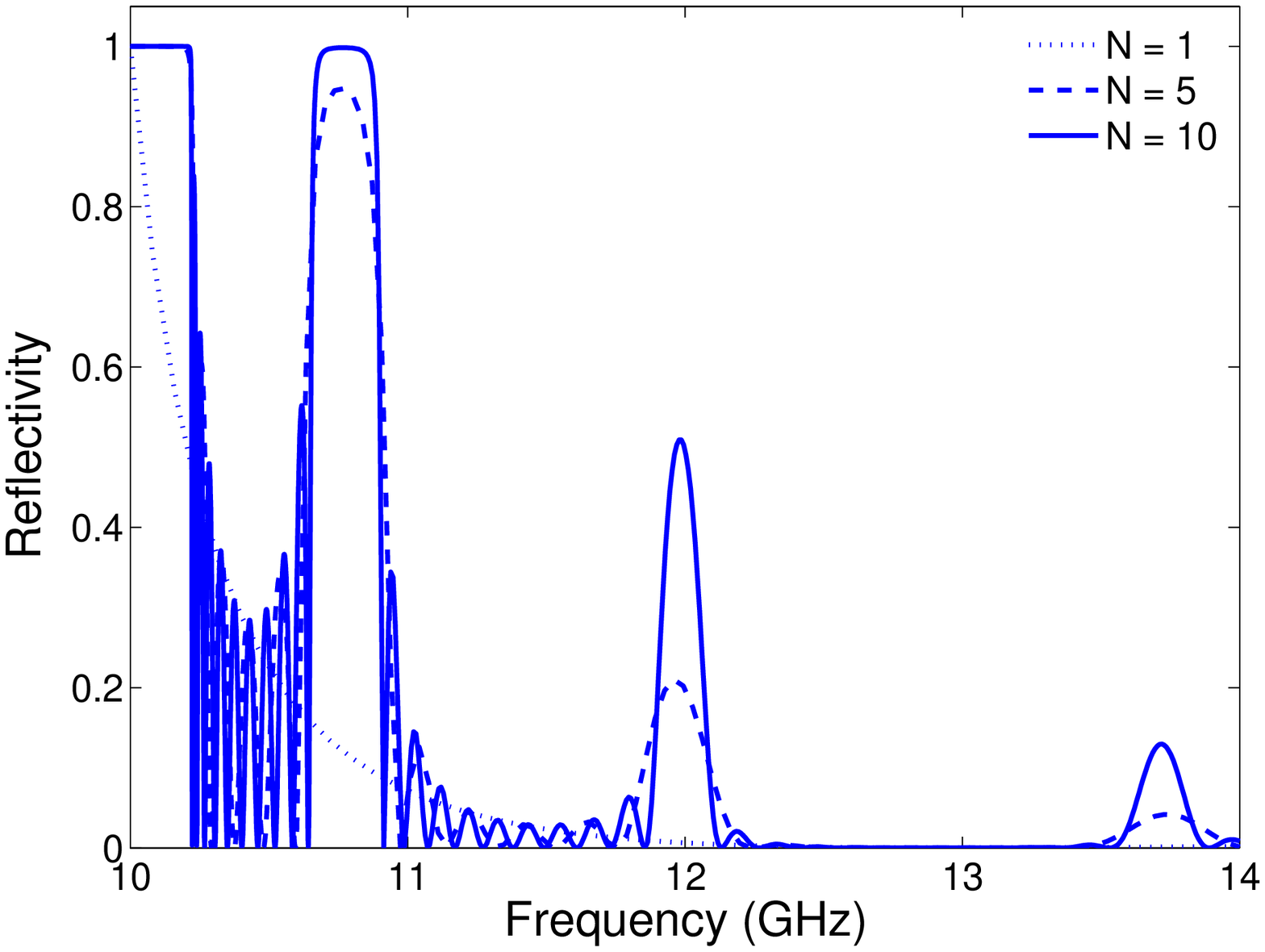}
\end{minipage}
\caption{Evolution of the magnonic band structure. The reflectivity of a finite magnonic crystal, with $N$ repeats of the unit cell, is plotted as a function of the repeat number. There is a DW nucleated in region 2 of the unit cell (left). For comparison, the result of an identical MC without the DW in the unit cell is also shown (right). The unit cell has $d_1 = 0.3 \mu$m and $d_2 = 0.1 \mu$m, and the anisotropy field in region 2 is reduced to 0.9 of that in region 1, $K_2 = 0.9 K_1$. Other parameter are identical to those used in Fig. \ref{sym-band}.}
\label{reflectivity}
\end{figure}

Partial band gap formation at the energy of an incident SW can be observed as a increased reflectivity. For a complete band gap, the reflectivity should be unity. Hence the SW reflectivity can serve as an indicator of band gap formation. The SW reflectivity of the DWMC, or any 1D structures composed of single domains or 2$\pi$ DWs, can be calculated using the method of propagation matrix \cite{Qiu00}. For this purpose, we need to know only two kinds of matrices: one is the propagation matrices describing the propagation of SW fields inside a medium, and the other is the interface matrices correlating the fields on both sides of an interface separating two media. For the case considered here, the propagation matrix in the single domain region is
$$P = \left( \begin{array}{cc} e^ {i k_1 d_1} & 0 \\ 0  &  e^ {- i k_1 d_1} \end{array} \right)$$
for the propagation of the field $\psi = (a_1 \exp (i k_1 y), b_1 \exp (- i k_1 y)) ^T$ from $y = - d_1$ to $y = 0$. A similar expression holds for the propagation matrix in the presence of the 2$\pi$ DW,
$$\tilde {P} = \left( \begin{array}{cc} e^ {i k_2 d_2} & 0 \\ 0  &  e^ {- i k_2 d_2} \end{array} \right),$$
describing the phase accumulated when the field $\tilde {\psi} = (a_2 \phi, b_2 \phi^*) ^T$ moves from $y = 0$ to $y = d_2$. The interface matrix at $y = - d_1$ is defined through the continuity equation $A \psi (-d_1) = \tilde {A} \tilde {\psi} (- d_1)$, which gives
$$A = \left( \begin{array}{cc} 1 & 1 \\ i k_1  &  - i k_1 \end{array} \right), \tilde {A} = \left( \begin{array}{cc} 1 & 1 \\ i q  &  - i q \end{array} \right)$$
With those interface and propagation matrices, the out-going wavefunction $\psi_R$ can be related to the incoming wavefunction $\psi_L$ by the transfer matrix $T = A^{-1} \tilde{M} (M \tilde{M})^N A$ through the relation $\psi_R = T \psi_L$. Matrices $M = A P A^{-1}$ and $\tilde{M} = \tilde{A} \tilde{P} \tilde{A}^{-1}$ are related to the interface and propagation matrices, and $N$ is the number of the repeats of the unit cell. Since we inject and detect SWs in regions with the same parameters as region 1, $N$ can actually be enumatated by the repeats of the DW. Fig. \ref{reflectivity} shows the reflectivity for three values of the number of unit cells, $N$ = 1, 5 and 10. It can be seen that with only $N$ = 5 unit cells, the lowest two bandgaps are already well developed. An increase to $N$ = 10 only improves the bandgap reflectivity slightly. Note that in Fig. \ref{reflectivity}, the bandgap development appears to be faster for the DWMC, as compared to the same conventional MC without the DW in the unit cell. However, this does not mean that DWMCs are superior to conventional MCs on this respect. The difference between those two MCs is caused by the different SW cutoff frequencies. When there is the DW in the unit cell, the cutoff frequency is increased significantly, to cover almost all the four bandgaps shown in Fig. \ref{reflectivity}, due to the small $d_2$. The only observation is that the bandgaps close to the cutoff frequency develops faster, simply because of their evanescent characteristics. For experimental realization of a MC, our calculation demonstrates that bandgaps will develop in the presence of the order of 10 unit cells. This conclusion is valid only for ideal interfaces. If interface roughness is inevitably present, either due to ion implantation or imperfect growth of materials, more unit cells may be needed to observe significant reflection of SWs within bandgaps.

\section{Discussion}
\label{discussion}
For the implementation of DWMCs, the nucleation or injection of DWs into the MC is the main obstacle. Although $\pi$ DWs are extensively studied due to their potential use as information carriers, 2$\pi$ DWs receive little attention, although they frequently appear during the demagnetization process of a magnet \cite{Hubert}. During the demagnetization process of a magnetic tunnel junction, 2$\pi$ DWs were observed using  Lorentz transmission electron microscopy \cite{Portier}. In magnetic nanorings, meta-stable 2$\pi$ DWs were observed by magnetic force microscopy \cite{Castano}. Later, the existence and stability of 2$\pi$ DWs was theoretically proven \cite{Muratov}. However, those randomly nucleated 2$\pi$ DWs are not amenable to be used in DWMCs, due to the difficulty in manipulating them. On this respect, multiple 2$\pi$ DWs can be injected into nanowires by cycling the polarity of field \cite{Jang}, or into a wedge shaped stripe by rotating field method \cite{Diegel}. In addition, Ar ion-implantation can be employed to facilitate the formation of multiple 2$\pi$ DW state \cite{Liedke}. Making use of the Oersted field generated by a current carrying wire, multiple 2$\pi$ DWs can also be nucleated \cite{Thomas12,GonzalezOyarce-apl}. For the manipulation of DW chirality, recent micromagnetic studies showed that gold shunt pads could be employed to select the chirality of 2$\pi$ DWs \cite{Zhang}. For additional control of the DW position besides of the ion-implantation method, triangular antinotches can be used to pin the location of 2$\pi$ DWs \cite{GonzalezOyarce-prb}, similar to the case of $\pi$ DWs.

To create periodic DW structures, we can use ion-implantation to reduce the anisotropy in region 2 to create a potential well for DWs to settle in, and through cycling field polarity or rotating field method to nucleate and inject 2$\pi$ DWs. With this method, it is easier to inject $\pi$ DWs in ion-implanted magnetic structures to fabricate $\pi$ DWMCs. Laser local heating can be used to enhance the probability of locally nucleating and traping DWs \cite{Tetienne}. In superlattices with anti-ferromagnetic or ferromagnetic coupling between adjacent magnetic layers, layers with lower DW energy can also accommodate DWs, following proper sequences of field preparation. But this would require a very large thickness of the soft layers. Periodic 2$\pi$ DW structure was observed in double-layer Fe nanowires on W(110) during the demagnetization process \cite{Kubetzka03}, which would be an ideal model system for the study of DWMCs. Finally, by applying a large hard axis field, and then reduce it to zero, 2$\pi$ stripe DWs can appear \cite{Leaf06}. This could be the simplest method to create DWMCs.
\section{Conclusion}
In summary, the SW dispersion relation of a 2$\pi$ DW and the corresponding magnonic band structure of a 1D DWMC have been analytically calculated. Due to the continuous magnetization rotation of the DW, the effective potential in the eigenequation for SWs is periodic, with only half of the period of the magnetization profile. The consequence of the periodic potential is to shift the energy minima of the SW spectrum to the Brillouin zone boundaries. Despite of the periodic potential experienced by SWs, there is no band gap opening in the whole Brillouin zone, which is caused by the absence of back scattering of SWs provided by time-reversal symmetry. For the SW eigenfunction, the 2$\pi$ DW induces a Berry phase, in additional to the dynamical phase related to the crystal momentum. Given those unique features of SWs propagating in the 2$\pi$ DW, the DWMC exhibits a different magnonic band structure as compared to a reference MC containing only uniformly magnetized domains, which is made possible by the continuous magnetization rotation of the DW. We investigate systematically the band structure evolution as a function of the anisotropy modulation and unit cell size. It is found that the magnonic band gaps can close, when magnons form quantum well states in the unit cell. Due to the appearance of two phases, the dynamic phase and the Berry phase, one advantage of including DWs in the unit cell is to obtain an additional control over the band structure of DWMCs. If realizable, reconfigurable switching between band structures of DWMCs and domain MCs can be achieved through the nucleation and annihilation of DWs by application of a magnetic field.

\section*{Acknowledgements}
This work was financially supported by the National Natural Science Foundation of China (No. 11374373) and the Natural Science Foundation of Hunan Province (No. 13JJ2004).

\end{document}